\documentclass[aps,showpacs,pra,twocolumn,superscriptaddress]{revtex4}
\usepackage{epsfig,amssymb,amscd}
\usepackage{psfrag}

\newcommand{\ket}[1]{|#1\rangle}
\newcommand{\yb}{$^{171}$Yb$^+$ }
\newcommand{\half}{\ensuremath{{\textstyle \frac{1}{2}}}}

\newcommand{\third}{\ensuremath{{\textstyle \frac{1}{3}}}}
\newcommand{\Id}{1\hspace{-0.56ex}{\rm I}}

\begin{document}
\bibliographystyle{prsty}

\title{Trapped-ion qutrit spin molecule quantum computer}
\author{D. Mc Hugh}
\affiliation{Department of Mathematical Physics,
National University of Ireland Maynooth, Maynooth, Co. Kildare, Ireland}
\author{J. Twamley}\email{Jason.Twamley@nuim.ie}
\affiliation{Dept. of Mathematical Physics, National University of Ireland, Maynooth,  Co. Kildare, Ireland}
\affiliation{Centre for Quantum Computer Technology, Macquarie University, Sydney, New South Wales 2109, Australia}

\received{\today}

\begin{abstract}
We present a qutrit quantum computer design using trapped ions in the presence 
of a magnetic field gradient. The magnetic field gradient induces a "spin-spin"
type coupling, similar to the \mbox{J-coupling} observed in molecules, 
between the qutrits which allows conditional quantum logic to take place. We describe in some detail, how one can execute specific one and two qutrit quantum gates, required for universal qutrit quantum computing.
\end{abstract}
\vskip 0.1cm
\maketitle

Quantum information and quantum computing \cite{books}, have 
made huge advances both theoretically and experimentally in recent times. There
are many different proposals 
\cite{cz,nmr,qdots,kane,fullerene,cqed,squid,helium,klm,anyons},
for the physical implementation of a quantum 
computer, all of which are specified by the physics of their qubit systems and the nature of the interactions between
qubits. The latter is necessary for the execution of conditional qubit logic, a requirement for universal qubit quantum computation. The qubit is of course
the obvious unit of a quantum computer given our classical computer historical dependance on binary logic. 
Quantum algorithms and protocols for quantum communication and cryptography
have been studied extensively with the qubit as the information storage and 
transport medium. Recently $d$-level quantum systems, or qudits, have started
to be considered seriously in terms of generalizing and improving qubit-based quantum algorithms and protocols. 
Interesting results have emerged in the qutrit ($d=3$) case.
It has been shown that quantum cryptography protocols are more robust against
eavesdropping attacks when qutrits are used \cite{bruss,cerf,durt}. 
Quantum bit commitment and coin-flipping protocols are more secure
with entangled qutrits  than with qubits \cite{spekkens}. 
Indeed, it is expected that qutrit-based quantum information processing 
will be more powerful
than other qudit implementations since they optimize the 
Hilbert space dimensionality  \cite{clark}. 
More speculative is the possibility of a higher error-tolerance for fault-tolerant qutrit quantum computation \cite{knill}. 
However, there have not been many qudit-based quantum computer proposals. 
As far as we are aware there has been only one qutrit quantum computer 
proposal using trapped ions  which generalises
the original Cirac-Zoller design \cite{klimov}.

Here we describe a modification to previous work \cite{wundprl,wundlaser,dmchjt},
on ion trap quantum computers where now qutrits store
the quantum information. An axial magnetic field gradient is applied across an ion chain
that allows the three hyperfine Zeeman energy levels of each ion, forming the qutrit, to
be individually frequency addressed. It also introduces an inter-qutrit
coupling that facilitates conditional quantum logic between qutrits. Previously
the operation of an ion trap quantum computer in the presence of a magnetic
field gradient has been discussed with qubits as the unit of quantum information.
In \cite{wundprl,wundlaser}, it is shown that all quantum gate operations, 
normally requiring optical irradiation, can be implemented using long
wavelength radiation due to the effects of the magnetic field gradient and trapping potential. 
The gradient also introduces
a term in the Hamiltonian that is analogous to the spin-spin coupling observed
between nuclei in molecules in NMR. This coupling can be used to perform quantum logic.
This idea is investigated further for ions in a linear array of microtraps 
\cite{dmchjt}. In the only other qutrit ion trap quantum computer proposal 
\cite{klimov}, the quantized collective vibrational motion of the linearly
trapped ions is used as a quantum bus to preform quantum logic.

\begin{figure}[p]
\psfrag{a}{{\large $\ket{2}$}}
\psfrag{b}{{\large$\ket{1}$}}
\psfrag{c}{{\large$\ket{0}$}}
\psfrag{d}{{\large$\omega_{01}$}}
\psfrag{e}{{\large$\omega_{12}$}}
\begin{center}
\setlength{\unitlength}{1cm}
\begin{picture}(6,10)
\put(.5,.5){\includegraphics[width=4cm]{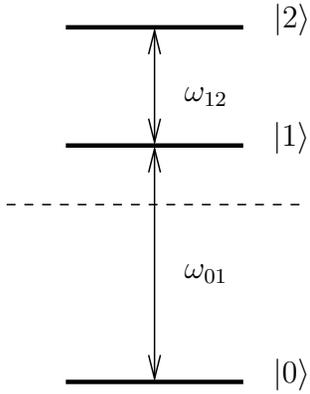}}
\end{picture}
\end{center}
\caption{The qutrit levels $\{\ket{0},\ket{1},\ket{2}\}$ under consideration
with $\omega_{12}$ and $\omega_{01}$ the resonant frequencies for
the $\ket{1}\leftrightarrow\ket{2}$ and $\ket{0}\leftrightarrow\ket{1}$ 
transitions respectively.}
\label{fig:qutrit}
\end{figure}

\begin{figure}[p]
\begin{center}
\setlength{\unitlength}{1cm}
\begin{picture}(6,10)
\put(-1,.5){\includegraphics[width=7cm]{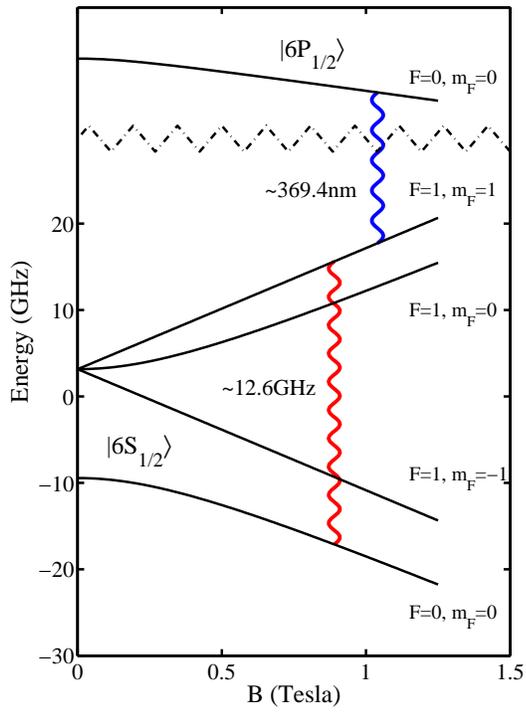}}
\end{picture}
\end{center}
\caption{The hyperfine Zeeman levels of an \yb ion in a spatially varying
magnetic field.}
\label{fig:yblevels}
\end{figure}

In this work we consider $N$ ions in a linear ion trap in the presence of a magnetic field
gradient. Three unequally spaced hyperfine Zeeman levels serve as our
qutrit (Fig \ref{fig:qutrit}). We will refer throughout to the \yb ion as an example 
qutrit, particularly the $F=1$ hyperfine Zeeman levels shown in Fig. 
\ref{fig:yblevels}. Our computational basis is now 
$\{\ket{0},\ket{1},\ket{2}\}$ written as $\ket{0}=(0,0,1)^{\top}$,
$\ket{1}=(0,1,0)^{\top}$ and $\ket{2}=(1,0,0)^{\top}$.
We denote by $\omega_{01}^{(n)}$ and $\omega_{12}^{(n)}$ the frequencies
resonant with the $\ket{0}\leftrightarrow\ket{1}$ and
$\ket{1}\leftrightarrow\ket{2}$ transitions for the $n$th ion and let
$\Delta_n=\omega_{01}^{(n)}+\omega_{12}^{(n)}$ and 
$\delta_n=\omega_{01}^{(n)}-\omega_{12}^{(n)}$. For each ion these 
frequencies, due to the spatial dependence of the magnetic field, are a
function of the ions' positions. The internal electronic Hamiltonian 
describing the spin degrees of freedom of a single ion is given by 
$H_{sp,n}=\half\hbar Z_n$ with $Z_n$ expressed in the above computational 
basis as 
\begin{eqnarray}
Z_{n}&=&\left( \begin{array}{ccc}
\Delta_n & 0 & 0 \\
0 & \delta_n & 0 \\
0 & 0 & -\Delta_n
\end{array} \right)\;\;.
\end{eqnarray}
As in \cite{wundprl, wundlaser}, the ions sit in an effectively 1d harmonic oscillator potential along the trap
axis and feel their
mutual Coulomb repulsion. Expanding this potential around their equilibrium
positions allows their vibrational motion to be treated collectively 
and the 
Hamiltonian for the motional degrees of freedom of the ions in normal 
coordinates is
\begin{eqnarray}
H_{vib}&=&\frac{1}{2m}\sum_nP^2_{Q,n}+\frac{m}{2}\sum_n\nu_n^2Q_n^2\;\;.
\end{eqnarray}
The local and normal coordinates are related by $q=D Q$, with $D$
the unitary transformation that diagonalises the Hessian, $A$, of the ions
potential evaluated at the equilibrium positions of the ions, $z_{0,\,n}$ [See refs \cite{wundlaser,sasura}].

The magnetic field gradient, $\frac{dB}{dz}\equiv b$, means the ions feel
a spatially varying magnetic field, $B(z)=B_0+b\cdot z$. This introduces a 
new term into the spin part of the Hamiltonian,
\begin{eqnarray}
H_{el,n}&=&\frac{1}{2}\hbar \left.Z_n\right|_{z_{0,n}}+\frac{\hbar}{2}\left.\frac{d Z_n}{d z}\right|_{z_{0,n}} q_n\;\;.
\end{eqnarray}
Setting $M_n=d Z_n/d z|_{z_{0,n}}$ for easier notation, as long as 
$<\half\hbar M_nq_n>$ is
much smaller than the ground state energy of the collective vibrational 
motion, then this new term will have a negligible effect on the normal modes
and can be treated as a perturbation. This places a limit on the size of the
magnetic field gradient but is no more stringent than other constraints 
discussed later. The resultant Hamiltonian is
\begin{eqnarray}
\nonumber
H&=&\frac{\hbar}{2}\sum_n Z_n+\frac{\hbar}{2}\sum_n\hat{M}_n\sum_l D_{ln}Q_l
\\\nonumber 
&&+\frac{1}{2m}\sum_nP_{Q,n}^2 +\frac{m}{2}\sum_n\nu_n^2Q_n^2\;\;,\\
\nonumber &=&\frac{\hbar}{2}\sum_nZ_n+\frac{1}{2m}\sum_lP_{Q,n}^2\\
\nonumber &&+\frac{m}{2}\sum_l\nu_l^2
\left(Q_l+\frac{\hbar}{2m\nu_l^2}\sum_nM_n D_{ln}\right)^2\\
&&-\frac{m}{2}\sum_l\nu_l^2\left(\frac{\hbar}{m\nu_l^2}\sum_n M_nD_{ln}\right)^2\;\;.
\end{eqnarray}
We move to a rotating frame where the spin and motional degrees of freedom
are decoupled via the unitary transformation 
$U=\exp\left[i\left(\frac{\hbar}{2m\nu_l^2}\sum_nM_nD_{ln}\right)P_{Q,l}\right]$,
yielding $\tilde{H}=UHU^{\dagger}$, with
\begin{equation}
\tilde{H}=\frac{\hbar}{2}\sum_nZ_n+ \sum_n\hbar\nu_na_n^{\dagger}a_n-H_{MM}\;\;,
\label{ham}
\end{equation}
where $H_{MM}=\frac{m}{2}\sum_l\nu_l^2\left(\frac{\hbar}{m\nu_l^2}\sum_nM_nD_{ln}\right)^2$,
and the position and momentum operators have been expressed in terms of
creation and annihilation operators in the usual way. 
$H_{MM}$ can be expressed as $H_{MM}=\frac{\hbar}{2}\sum_{n<m}J_{nm}M_nM_m$
where \begin{equation} 
J_{nm}=\frac{\hbar}{2m}\sum_l\frac{1}{\nu_l^2}D_{ln}D_{lm}\;\;. \label{J} 
\end{equation}
The Hamiltonian $\tilde{H}$ in (\ref{ham}), describes $N$ individually addressable 
qutrits coupled through $H_{MM}$, a 
``spin-spin'' type interaction which we will presently show can be used to 
perform conditional 
quantum logic between qutrits. There are no extra experimental requirements
compared to the setup proposed in \cite{wundprl,wundlaser}.

It is necessary to demonstrate how single qutrit operations and conditional
logic, the most basic ingredients of a universal quantum computation, can be implemented in
this modified design.

In order to perform single qutrit operations, we use the fact that the two transitions 
$\ket{0}\leftrightarrow\ket{1}$, and $\ket{1}\leftrightarrow\ket{2}$, have 
different resonant frequencies allowing the operations $U_{01}$, and  $U_{12}$,
\begin{eqnarray}
\nonumber
U_{12}(\theta,\phi)&=&\left( \begin{array}{ccc}
\cos\frac{\theta}{2} & ie^{i\phi}\sin\frac{\theta}{2} & 0 \\
ie^{-i\phi}\sin\frac{\theta}{2} & \cos\frac{\theta}{2} & 0 \\
0 & 0 & 1  
\end{array} \right)\;\;,\\
U_{01}(\theta,\phi)&=&\left( \begin{array}{ccc}
1 & 0 & 0 \\
0 & \cos\frac{\theta}{2} & ie^{i\phi}\sin\frac{\theta}{2} \\
0 & ie^{-i\phi}\sin\frac{\theta}{2} & \cos\frac{\theta}{2}  
\end{array} \right)\;\;,
\label{units}
\end{eqnarray}
to be carried out. Since the $\ket{0}\leftrightarrow\ket{2}$ transition is 
forbidden, any rotation between these two states 
requires a $\pi$-pulse between $\ket{0}$ and
$\ket{1}$, $U_{01}(\frac{\pi}{2},0)$, the required gate on 
$\ket{1}\leftrightarrow\ket{2}$, $U_{12}(\theta,\phi)$, followed by 
another $\pi$-pulse on $\ket{0}\leftrightarrow\ket{1}$.
Also required is the unitary differential phase operation,
\begin{eqnarray}
U_D&=&\left( \begin{array}{ccc}
e^{i\rho} & 0 & 0\\
0 & e^{i\sigma} & 0\\
0 & 0 & e^{-i(\sigma+\rho)} \end{array} \right)\;\;,
\end{eqnarray}
where this gate is a composition of ``$\sigma_z$'' operations on the 
$\ket{0}\leftrightarrow\ket{2}$
and $\ket{0}\leftrightarrow\ket{1}$ transitions that are themselves 
compositions of $U_{12}$ and $U_{01}$ operations. 
In particular,
$U_D=(Z_{02})_{\rho}(Z_{01})_{\sigma}$ with 
$(Z_{ij})_{\rho}=H_{ij}U_{ij}(\rho,0)H_{ij}^{\dagger}$
and $H_{ij}=U_{ij}(\frac{\pi}{4},\frac{\pi}{2})$ is the Hadamard gate.
Thus the unitary operations given in (\ref{units}) allow us to generate 
any operation in $SU(3)$ \cite{arvind}. 

For gates between more than one qutrit, we propose to use the last part of
the Hamiltonian $\tilde{H}$ in (\ref{ham}). 
 We now consider in more detail the hyperfine Zeeman levels for the qutrits. For intermediate
magnetic field strengths $B$, such that $g_J\mu_BB\approx A$, where $g_J$ is the 
Land\'{e} 
g-factor, $\mu_B$, is the Bohr magneton and $A$, is the hyperfine constant, the
energy levels are described by the Rabi-Breit formula \cite{woodgate}. In this 
case $d\Delta_n/dz=g_J\mu_B b/\hbar$, and 
$d\delta_n/dz\approx g_J\mu_B b/\sqrt{2}\hbar$, and where $b$ is the constant gradient of the magnetic field $B(z)=B_0+bz$. We absorb the $g_J\mu_B b/\hbar$
factor into the definition of $J_{nm}$ and write $M=d Z/d z|_{z_0}$ 
in the matrix form
\begin{eqnarray}
M&=&\left( \begin{array}{ccc}
1 & 0 & 0 \\
0 & \frac{1}{\sqrt{2}} & 0 \\
0 & 0 & -1
\end{array}\right)
\label{M}
\end{eqnarray}
with now 
$J_{nm}=\frac{(g_J\mu_Bb)^2}{2\hbar}(A^{-1})_{nm}$ \cite{dmchjt}.
Expressing the operator $M$ in (\ref{M}) in terms of the generators of $SU(3)$ we have
\begin{equation} M=a_0\Id+a_3\lambda_3+a_8\lambda_8\;\;, \label{Msu3}\end{equation} where
\begin{equation} \begin{array}{cc}\lambda_3=\left( \begin{array}{ccc}
1 & 0 & 0 \\
0 & -1 & 0 \\
0 & 0 & 0
\end{array}\right)\;\;,& \lambda_8=\frac{1}{\sqrt{3}}\left(\begin{array}{ccc}
1 & 0 & 0 \\
0 & 1 & 0 \\
0 & 0 & -2
\end{array}\right) \end{array}\end{equation}
with $a_0=\frac{1}{3\sqrt{2}}$, $a_3=\frac{1}{2\sqrt{2}}(\sqrt{2}-1)$,
and $a_8=\frac{1}{2\sqrt{6}}(1+3\sqrt{2})$. $H_{MM}$ is thus a two-body
N-qutrit Hamiltonian as defined in \cite{nielsen}, and so, as arbitrary local
unitaries are possible, universal quantum computation can be performed.
More specifically, it has been pointed out in \cite{klimov} that a generalized 
XOR gate between qutrits, 
\begin{eqnarray}
{\mbox XOR}_{mn}\ket{j}_m\ket{k}_n=\ket{j}_m\ket{j\oplus k}\;\;,
\end{eqnarray}
where the ``$\oplus$'' operation now indicates addition modulo 3, can be 
decomposed into three operations, 
\begin{equation}
{\mbox XOR}_{mn}=F_mP_{mn}F_m^{-1}\;\;.
\end{equation}
The generalized Fourier transform, $F$, for one qutrit is defined by
\begin{eqnarray}
F\ket{j}&=&\frac{1}{\sqrt{3}}\sum_{l=0}^2e^{2\pi ilj/3}\ket{l}\;\;,
\end{eqnarray}
where $j=0,1,2$. 

The phase gate for qubits, 
$P_{qub}\ket{j}_m\ket{k}_n=\exp(i\pi jk)\ket{j}_m\ket{k}_n$, is 
sandwiched by two Hadamard gates on the target qubit to give the 
controlled-NOT or XOR operation.
The generalization of this gate to qutrits is $P_{mn}$. It is 
completely specified by
\begin{eqnarray}
P_{mn}\ket{j}_m\ket{k}_n=\exp (2i\pi jk/3)\ket{j}_m\ket{k}_n\;\;.
\label{P}
\end{eqnarray}
We can generate this gate between 2 qutrits using a 
combination of single qutrit rotations and evolution
under the two qutrit Hamiltonian $H_{12}=2\pi JM_1\otimes M_2$.
One pulse sequence for $P_{mn}$ is
\begin{eqnarray}
\nonumber
&&\left(Z_{01}^{(1)}\right)_{\alpha_1}\left(Z_{12}^{(1)}\right)_{\alpha_2}
\left(Z_{01}^{(2)}\right)_{\alpha_3}\left(Z_{12}^{(2)}\right)_{\alpha_4}\\
\nonumber &\cdots& \left(MM\right)_{\alpha_5}\left(X_{01}^{(1)}\right)
\left(MM\right)_{\alpha_6}
\left(X_{01}^{(1)}\right)\\
&\cdots& \left(X_{12}^{(1)}\right)
\left(MM\right)_{\alpha_7}\left(X_{12}^{(1)}\right)
\left(X_{12}^{(2)}\right)\\
\label{pulsephase}
\nonumber &\cdots& \left(MM\right)_{\alpha_8}\left(X_{12}^{(1)}\right)
\left(MM\right)_{\alpha_9}
\left(X_{12}^{(1)}\right)\left(X_{12}^{(2)}\right)
\end{eqnarray}
where $(X^{(n)}_{kl})$ represents a $\pi$-pulse on the 
$\ket{k}\leftrightarrow\ket{l}$ transition of the $n$th ion
i.e. $(X^{(n)}_{kl})=U_{kl}(\frac{\pi}{2},0)$ and 
$(MM)_{\theta}$ represents a period of evolution under the Hamiltonian 
$H_{12}$ for a time $t$ such that $2\pi Jt=\theta$.
While this is not a unique decomposition, the 
number of periods of evolution under $H_{MM}$ is minimal.
The angles $\{\alpha_i\}_{i=1}^9$ have been numerically determined and are 
shown in Table \ref{tab:alpha}.

Refocussing techniques \cite{books} developed for NMR quantum computing are
necessary here given the ``always-on'' nature of $H_{MM}$. For
qubits the interaction is $H_{SS}=2\pi J\sigma_z^{(1)}\sigma_z^{(2)}$. The
relations $\sigma_a\sigma_z\sigma_a=-\sigma_z$ for $a=x,y$, where 
$\{\sigma_i\}_{i=x,y,z}$ are the Pauli operators, are used to reverse the 
evolution under $H_{SS}$ \begin{equation} 
e^{itH_{SS}} = e^{-i\frac{\pi}{2}\sigma_x^{(1)}}e^{-itH_{SS}}
e^{-i\frac{\pi}{2}\sigma_x^{(1)}}.\end{equation}
Essentially the diagonal elements of $\sigma_z$ are permuted by the two
$\pi$-pulses. Combined with another period of evolution under $H_{SS}$, 
the trace-less property of $\sigma_z$ is exploited so that nothing
happens. In fact, all qubits coupled to the first qubit are refocussed by this.

For qutrits the de-coupling procedure is basically the same.
The spin-spin term $H_{MM}=2\pi JM_1M_2$ is composed of nine terms,
\begin{equation} 
M_1M_2= a_0\Id\otimes M+(a_3\lambda_3+a_8\lambda_8)\otimes M  \label{MMl}
\end{equation}
By applying pulses which simultaneously permute the entries of $\lambda_3$, and $\lambda_8$, over
three different permutations, in a manner similar to the qubit case, the combined evolution under the second
term in (\ref{MMl}) is removed. The sequence to refocus the $MM$ evolution over an arbitrary duration $T$, requires one to subdivide the duration into three equal-duration sub-evolutions $(MM)_\theta$, ($2\pi J T/3=\theta$). One can then obtain, 
\begin{equation} e^{i\phi}U_1U_2U_3R_1^{(2)}R_2^{(2)}=\Id \label{refocus}
\end{equation}
where \begin{eqnarray}
\nonumber U_1&=&(MM)_{\theta}\\
\nonumber U_2&=&X^{(1)}_{01}X^{(1)}_{12}(MM)_{\theta}X^{(1)}_{12}X^{(1)}_{01}\\
U_2&=&X^{(1)}_{12}X^{(1)}_{01}(MM)_{\theta}X^{(1)}_{01}X^{(1)}_{12}
\end{eqnarray} and \begin{eqnarray}
\nonumber R_1^{(2)}&=&\exp(3a_0a_3\theta) \\
R_2^{(2)}&=&\exp(3a_0a_8\theta).
\end{eqnarray}
The global phase factor, $\phi=3a_0^2\theta$, is irrelevant while the 
two single-qutrit pulses on the second qutrit are 
due to the first term in (\ref{MMl}) and can be easily reversed. 
The three periods of evolution required are expected since the matrices
are 3-dimensional.

Readout of the final state of the qutrit register takes two steps.
The entire ion chain is illuminated with optical radiation and the
observed fluorescence spatially resolved. The radiation frequency is chosen
so that if the qutrits are projected onto $\ket{2}$ this is then detected.
A $\pi$-pulse is then applied on the
$\ket{1}\leftrightarrow\ket{2}$ transition of each ion and the ion string
illuminated
again. Fluorescence now indicates projection onto $\ket{1}$ while its absence
means the qutrit is in $\ket{0}$.

We now an describe explicit example using the ${}^{171}$Yb$^+$ ion. 
The hyperfine constant for \yb is $A=2\pi\ 12.6$GHz. In a magnetic field its 
levels are split as shown in Fig.
\ref{fig:yblevels}. In fields of around 0.45T, the Rabi-Breit region, the
good quantum numbers are $F$ and $M_F$ and our logical states are
$\ket{0}=\ket{{\mbox 6S}_{\half}F=1,M_F=-1}$,
$\ket{1}=\ket{{\mbox 6S}_{\half}F=1,M_F=0}$, and
$\ket{2}=\ket{{\mbox 6S}_{\half}F=1,M_F=1}$.
The transition frequencies $\omega_{01}$, and $\omega_{12}$, are about
$3.7$GHz and $8.9$GHz, respectively while the differences between the resonance 
frequencies of two neighbouring ions in a trap of frequency $2\pi$ 200kHz
and a magnetic field gradient of $b=100$T/m are $\delta\omega_{01}\approx11$
MHz, and $\delta\omega_{12}\approx2$MHz. The operator $M_n$ in (\ref{M}) has an
element which is approximately 1/$\sqrt{2}$. This approximation introduces a 
constraint on the size of the magnetic field gradient if we require $M_n$ to be
constant over the ion chain within an accepted error $\epsilon_M$.
Numerical calculations in \cite{james,steane}, give the minimum distance between
ions in a chain as $\Delta z_{min}(N)=2.018\gamma N^{0.559}$ where
$\gamma=(q^2/4\pi\epsilon_0m\nu_1^2)^\third$. As a conservative estimate, let us
say that the ions are equally spaced at $\delta z=1.5\Delta z_{min}(N)$. This
implies that $b<0.03N^{-0.441}\nu_1^{\frac{2}{3}}$, limiting the size 
of the
magnetic field gradient. On the other hand we need to be able to frequency
discriminate between qutrit transitions on neighbouring ions, and be far
enough away that no vibrational motion will be excited i.e.
$(d\omega_{01}/d z)\delta z\le 2\nu_N+\nu_1$. Using the
numerical result \cite{wundprl} that $\nu_N=(2.7+0.5N)\nu_1$, this imposes
a minimum size on the magnetic field gradient of
$b\ge 1.5\times10^{-9}\nu_1^{\frac{5}{3}}(3.2N^{0.559}+0.5N^{1.559})$.
For an axial trap frequency of $\nu_1=2\pi\ 200$kHz containing 10 ions and
$\epsilon_M=0.01$
the magnetic field gradient is limited by $30{\rm T/m}\le b\le200{\rm T/m}$.
These limits also determine the maximum number of
ions we can place in the trap, assuming other considerations related to the
ratio of the axial and radial trapping frequencies are satisfied \cite{sasura}.
For an axial trap frequency of $\nu_1=2\pi\ 200$kHz, the maximum
number of ions is about 30 satisfying the above limits when the magnetic field 
gradient is 120T/m.
The expression for the J-couplings in (\ref{J}) is the same as that in the
qubit case. For 10 ${}^{171}$Yb${}^+$ ions at a trap frequency
of $2\pi$ 200kHz and magnetic field gradient 120T/m gives a nearest-neighbour
coupling of about $J=2\pi\ 1.2$kHz.

In summary, we have presented a modification of previous designs for ion quantum computation with magnetic field gradients but where the quantum information
is now manipulated and stored in qutrits. A magnetic field gradient allows for individual 
qutrit addressing  and introduces a qutrit-qutrit coupling for quantum logic.
The scheme requires no additional physical resources.

{\bf Acknowledgement}\\
The authors gratefully acknowledge Christof Wunderlich for helpful comments
on this manuscript.
D. McHugh kindly acknowledges support from Enterprise-Ireland Basic Research
Grant SC/1999/080. The work was also supported by the EC IST FET project 
QIPDDF-ROSES IST-2001-37150 and Science Foundation Ireland.


\begin{table}[ht]
\begin{tabular}{|c|c|}
\hline
$\alpha_1$&-0.5628\\
\hline
$\alpha_2$&-0.2604\\
\hline
$\alpha_3$&-1.9045\\
\hline
$\alpha_4$&-2.4299\\
\hline
$\alpha_5$&-16.5854\\
\hline
$\alpha_6$&19.1630\\
\hline
$\alpha_7$&-0.2738\\
\hline
$\alpha_8$&5.3918\\
\hline
$\alpha_9$&0.3045\\
\hline
\end{tabular}
\caption{
The angles $\alpha_i$ written in multiples of $\pi$ required
in order to execute the pulse sequence for the phase gate for qutrits given
in (\ref{pulsephase}).}
\label{tab:alpha}
\end{table}

\end{document}